# Time-Division Multiplexing Light Field Display with Learned Coded Aperture

Chun-Hao Chao[*], Chang-Le Liu[*], and Homer H. Chen, *Fellow, IEEE*

*Abstract*—Conventional stereoscopic displays suffer from vergence-accommodation conflict and cause visual fatigue. Integral-imaging-based displays resolve the problem by directly projecting the sub-aperture views of a light field into the eyes using a microlens array or a similar structure. However, such displays have an inherent trade-off between angular and spatial resolutions. In this paper, we propose a novel coded time-division multiplexing technique that projects encoded sub-aperture views to the eyes of a viewer with correct cues for vergence-accommodation reflex. Given sparse light field sub-aperture views, our pipeline can provide a perception of high-resolution refocused images with minimal aliasing by jointly optimizing the sub-aperture views for display and the coded aperture pattern. This is achieved via deep learning in an end-to-end fashion by simulating light transport and image formation with Fourier optics. To our knowledge, this work is among the first that optimize the light field display pipeline with deep learning. We verify our idea with objective image quality metrics (PSNR, SSIM, and LPIPS) and perform an extensive study on various customizable design variables in our display pipeline. Experimental results show that light fields displayed using the proposed technique indeed have higher quality than that of baseline display designs.

## I. Introduction

TRADITIONAL stereoscopic displays offer 3-D depth perception by presenting two images with disparity. Although the binocular disparity is a dominant cue that drives the eye vergence for depth perception, stereoscopic displays cannot provide correct accommodation cue that matches the eye vergence because the images are projected at a fixed focal distance. Any such mismatch (aka vergence-accommodation conflict, VAC) can easily cause eye strain, headache, or even loss of depth perception because human eyes have evolved over the history of mankind to have consistent accommodation and vergence.

To address the issue, several kinds of 3-D displays have been proposed, such as varifocal displays [1]–[3], multifocal displays [4]–[12], holographic displays [13]–[15]. However, varifocal displays and multifocal displays both require additional elements such as eye trackers and screen stacks, while holographic displays require high-resolution spatial light modulators. Integral-imaging-based light field displays (II-based LFDs) [16]–[24] provide correct cues for vergence-accommodation reflex without such extra elements. 3-D perception is provided by, for example, projecting different views of a scene to the eye with a microlens array placed somewhere between the micro display panel and the eyepiece. However, there is a trade-off between spatial and angular resolutions, and severe undersampling in either spatial or angular domain may degrade image quality.

In this paper, we propose a novel idea called coded time-division multiplexing (CTDM) that leverages coded aperture imaging and time-multiplexing to resolve the undersampling issue for light field display (LFD). By rapidly changing the learned coded aperture pattern of the display optics, decoded images can be projected into the viewer's eye. Through the persistence of vision, these projected images would be perceived by the human eye as if they are summed together to form a refocused image.

The term "light field display" used here refers to a display system where the interval of viewpoints is made smaller than the pupil diameter of human eyes. Such a system allows users to freely focus at any depth of the scene. It should not be confused with displays that allow users to view a scene from different viewpoints [25], which do not solve the accommodation problem.

The contributions of our work are highlighted as follows:
- We propose a novel idea called CTDM for light field display that offers correct cues for vergence-accommodation reflex.
- Each component, such as the coded aperture pattern, in the CTDM-LFD pipeline is modeled with differentiable functions and optimized by deep learning. The optimized components greatly help generate high quality light field.
- The transformation performed by the proposed light field encoding network facilitates CTDM-LFD to generate light

[†]This work was supported in part by grants from the Ministry of Science and Technology of Taiwan under Contracts 110-2221-E-002-108-MY3, 110-2813-C-002-266-E, and 110-2218-E-002-032-MBK, in part by grants from National Taiwan University under Contract 110L880602, and in part by GPU computing resources from NVIDIA AI Technology Center (NVAITC). (*C.-H. Chao and C.-L. Liu are co-first authors.*) (*Corresponding author is H. H. Chen.*)

C.-H. Chao is with the Department of Electrical Engineering, National Taiwan University, Taipei 10617, Taiwan (e-mail: b06901104@ntu.edu.tw).

C.-L. Liu is with the Graduate Institute of Communication Engineering, National Taiwan University, Taipei 10617, Taiwan (e-mail: b05901017@ntu.edu.tw).

H. H. Chen is with the Department of Electrical Engineering, Graduate Institute of Communication Engineering, and Graduate Institute of Networking and Multimedia, National Taiwan University, Taipei 10617, Taiwan (e-mail: homer@ntu.edu.tw).

fields with better perceptual quality than that of the baseline TDM-LFD.

## II. RELATED WORK

Our work described in this paper is related to digital refocusing, coded aperture imaging, and 3-D display. This section reviews related work.

### A. Digital Refocusing

Digital refocusing, which allows a photograph to be focused at a different depth after taking the shot, has been widely investigated in image processing and computer graphics. As described in the seminal paper on light fields by Levoy et al. [26], digital refocusing can simply be implemented through the shift and add operations on light field sub-aperture views. Ng et al. [27] exploited the fact that a photograph is a 2D projection of a 4D light field and used the Fourier Slice Theorem to achieve digital refocusing. However, such refocusing algorithms assume a dense light field is readily available.

For digital refocusing from a sparse set of sub-aperture views, Bando et al. [28] used blur estimation and deconvolution to generate multiple deconvolved images of an input photograph and stitched the re-convolved images back together to synthesize a refocused image. Chang et al. [29] leveraged depth maps to compute different circles of confusion convolution kernels to create natural synthetic bokeh. Xiao et al. [30] leveraged view synthesis and aliasing detection to replace aliasing defocus regions with non-aliasing image patches. On the other hand, the advent of deep learning gave rise to learning-based refocusing. Dayan et al. [31] used a CNN to regress ground truth refocused images, and Sakurikar et al. [32] leveraged conditional generative adversarial networks to synthesize refocused images.

Existing digital refocusing methods involve either carefully crafted image operations such as blur estimation and deconvolution or large neural networks. These highly complex operations are hard to be integrated into near-eye light field displays, since these techniques cannot be realized using optics and that light field displays present a full light field rather than refocused images to the viewer. We are interested in a problem fundamentally different from these previous works on light field imaging. Our focus is in digital refocusing for near-eye light field displays, for which the refocusing operation is performed instantly by human eye.

### B. Coded Aperture Imaging

In the past, coded aperture was widely used for X-ray and gamma-ray imaging [33]–[37]. Because X-ray and gamma-ray are difficult to be refracted or reflected by common optical elements like lenses or mirrors, pinhole cameras are used for X-ray and gamma-ray detection. However, the tiny pinhole results in low light transmittance and low signal-to-noise ratio. To address the problem, scatter-hole [34] and coded-aperture cameras [33], [35]–[37] were developed, and specifically designed apertures were used to enhance photon collection and image quality.

Coded aperture has also been employed in computational imaging with visible light, like motion deblurring [38], depth estimation [39], high-dynamic-range imaging [40], and light field capturing [41]–[43]. However, the exploitation of coded aperture for light field displays is fairly new [44], [45].

TABLE I
NOMENCLATURE

| Symbol | Description |
|---|---|
| $L(x, y, s, t)$ | 4-D light field. The pairs $(x, y)$ and $(s, t)$ represent the spatial and angular coordinates, respectively. |
| $L_s(\mathbf{x})$ | Sub-aperture image at $\mathbf{s} = (s, t)$ or the view captured at the angular coordinates $(s, t)$. $\mathbf{x} = (x, y)$. |
| $\mathbf{I}_c$ | Encoded light field $\mathbf{I}_c = \{I_{c,1}(\mathbf{x}), I_{c,2}(\mathbf{x}), \ldots, I_{c,N}(\mathbf{x})\}$. |
| $P$ | Pupil function. |
| $\mathcal{F}$ | Non-unitary Fourier transform. We drop the constant coefficient for simplicity. |
| $\mathcal{F}^{-1}$ | Inverse Fourier transform. We drop the constant coefficient for simplicity. |

Just like coded apertures can be used for imaging, they can also be used in displays. The seminal work by Wetzstein et al. [45] leveraged coded apertures to increase the depth of field of projector systems. Inspired by this work and the coded aperture light field acquisition works [43], [46], [47], we design a light field display that simultaneously improves spatial and angular resolution of displayed light fields using time multiplexed coded aperture patterns.

### C. 3-D Display

Most traditional 3-D displays or stereoscopic displays present a pair of stereo images to a user. The user can sense 3-D objects with eye convergence; however, rendering the virtual images at a fixed depth, which most such displays do, may cause fatigue to the user because of VAC.

To address the VAC issue, various techniques such as varifocal displays [1]–[3], multifocal displays [4]–[12], holographic displays [13]–[15], volumetric displays [48], and light field displays (LFDs) [18]–[20], [22], [23], [49] have been proposed. A more detailed review of these techniques can be found in the survey by Zabels et al. [50]. Among the light field displays, II-based LFD provides 3-D information by rendering a number of views of the scene from different directions without complicated switchable elements or eye tracker required by varifocal and multifocal displays and without the spatial light modulator required by holographic displays. Although II-based LFD has been adopted in optical see-through head-mounted displays [22], [49], it has low spatial resolution (equal to the original display resolution divided by the number of lenses in the microlens array). This motivates us to explore the integration of coded aperture imaging with time-multiplexing by deep learning.

## III. SDM-LFD AND TDM-LFD

In this section, we explain how the time-division multiplexing light field display (TDM-LFD) overcomes the drawback of the space-division multiplexing light field display (SDM-LFD). The discussion here paves the way for the development of the proposed coded time-division multiplexing light field display (CTDM-LFD) in Section IV.

We consider the 4D light field representation proposed by Levoy and Hanrahan [26] and define the symbols used in this study in Table I. Note that, in practice, $s$ and $t$ are finite in most cases; therefore, without loss of generality, we assume each of them is bounded by *N*.

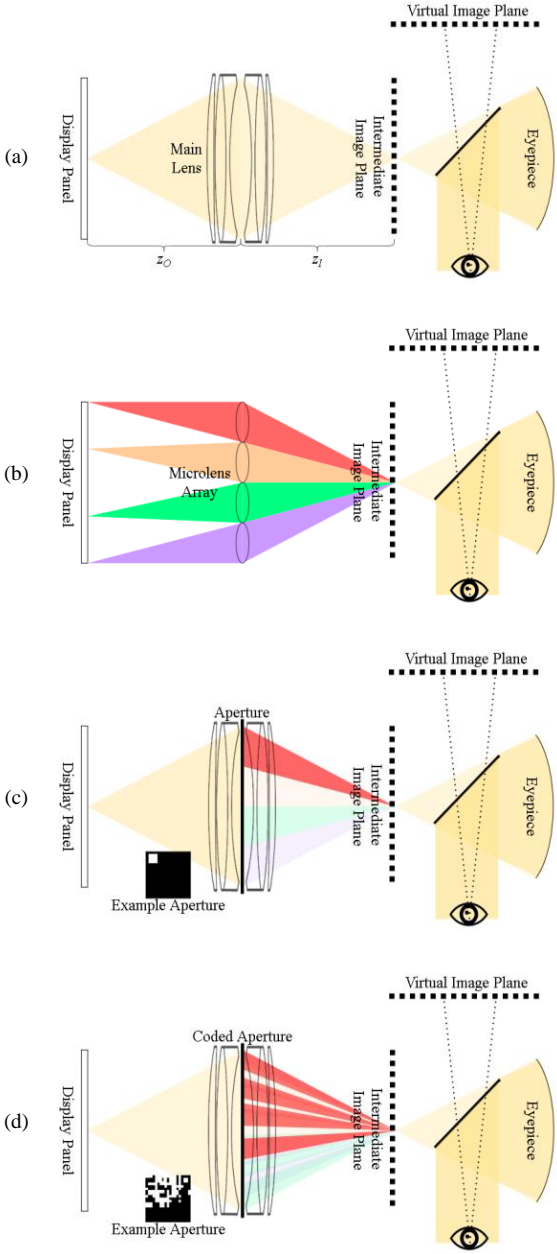

Fig. 1. The optical configuration of (a) Generic AR display, (b) SDM-LFD, (c) TDM-LFD, and (d) CTDM-LFD.

Most AR displays consist of a display panel, a lens system, and an eyepiece as the basic components. The light is emitted from the display panel to form an intermediate image at the intermediate image plane and the user perceives a virtual image through the eyepiece, as shown in Fig. 1(a). More specifically, the intermediate image is the collection of focus points formed by the main lens, and we call the plane where an intermediate image lies an intermediate image plane. The virtual image, on the other hand, is the collection of focus points formed by extensions of the diverging rays, and we call the plane where the virtual image lies a virtual image plane. We denote the object and image distances by $z_O$ and $z_I$, respectively. The eyepiece is an optical system that relays the intermediate images to the human eye.

### A. Space-Division Multiplexing Light Field Display

Similar to the traditional integral imaging-based light field camera [51], SDM-LFD uses a microlens array to divide the display panel into several sub-areas. Each sub-area of the panel corresponds to a sub-aperture view, and all sub-aperture views are displayed on the same intermediate image plane, as shown in Fig. 1(b). The perception of a light field is affected by projecting overlapping sub-aperture views to the human eye through an eyepiece. Although the design is straightforward, it requires high-precision manufacturing of the microlens arrays and has an inherent trade-off between the spatial and angular resolutions. For example, for a 3×3 light field display, the actual area that is perceived by the viewer is only one-ninth of the total area of the display panel. We resolve the drawback by using the time-division multiplexing technique and coded aperture pattern.

### B. Time-Division Multiplexing Light Field Display

For TDM-LFD, a switchable shutter is used to control the transport of light rays. The shutter is in synchronization with the display panel so that different sub-aperture views of a light field can be displayed in a time-multiplexing fashion, as shown in Fig. 1(c) (The only difference between Fig. 1(c) and 1(a) is that there is an additional switchable shutter in Fig. 1(c).). For example, when the sub-aperture view in the top left corner is displayed, the corresponding aperture (a small rectangular window) is turned on in the top left corner accordingly.

When a user focuses on a different virtual image plane, the user perceives a refocused image at the corresponding virtual image distance. By laws of image formation, this is equal to the shifting of the intermediate image plane. Because what the user perceives (virtual image) is conjugate with the intermediate image, we describe the refocusing process of the intermediate image by Fourier optics [52] in the following sections.

Consider first the formation of a refocused image from one input image. This corresponds to the case of a traditional AR display. Denote the complex field emitted from a monochromatic coherent point source with a wavelength $\lambda$ on the display panel and the complex field received on the intermediate image plane by $U_I(x,y)$ and $U_O(x,y)$, respectively. Using Fourier optics, we have

$$U_I(x_I, y_I) = \frac{1}{\lambda^2 z_I z_O} \iint_{-\infty}^{\infty} \iint_{-\infty}^{\infty} P(x_p, y_p)$$
$$\times \exp\left[j\frac{k}{2}\left(\frac{1}{z_I} + \frac{1}{z_O} - \frac{1}{F}\right)(x_p^2 + y_p^2)\right] \quad (1)$$
$$\times \exp\left[-jk\left(\left(\frac{x_O}{z_O} + \frac{x_I}{z_I}\right)x_p + \left(\frac{y_O}{z_O} + \frac{y_I}{z_I}\right)y_p\right)\right]$$
$$\times dx_p dy_p U_O(x_O, y_O) dx_O dy_O,$$

where $h$ is the impulse response of the lens system, $k=\frac{2\pi}{\lambda}$, and $z_O$ and $z_I$ are the object and image distances, respectively. Note that constant phase and quadratic phase factors have been dropped because they do not affect the image intensity, which is our primary interest [52].

TABLE II
OPTIMIZABLE COMPONENTS OF CTDM-LFD

| Symbol | Description |
|---|---|
| $f$ | A non-linear mapping from the original light field $L$ to the encoded light field that is input to the display panel. |
| $\hat{\mathbf{P}}$ | A set of coded apertures $\{\hat{P}_1, \hat{P}_2, ..., \hat{P}_k\}$. Input images are displayed and passed through the lens and coded apertures to form a light field. |

TABLE III
OPTIMIZABLE HYPER-PARAMETERS OF CTDM-LFD

| Symbol | Description |
|---|---|
| $n$ | The number of sub-aperture views sampled from the original light field. |
| $k$ | The number of images in the encoded light field, which is equal to the number of coded apertures in $\hat{\mathbf{P}}$. But $k$ is not necessarily equal to $n$. |
| $l$ | The spatial resolution of coded apertures. |

Letting $M = -z_I/z_O$ and $\tilde{x}_p = x_p/\lambda z_p$, we have

$$U_I(x_I, y_I) = \iint_{-\infty}^{\infty}\iint_{-\infty}^{\infty} P(\lambda z_i \tilde{x}_p, \lambda z_i \tilde{y}_p)$$
$$\times \exp\left[j\frac{k}{2}\left(\frac{1}{z_O} + \frac{1}{z_I} - \frac{1}{F}\right)((\lambda z_I \tilde{x}_p)^2 + (\lambda z_I \tilde{y}_p)^2)\right] \quad (2)$$
$$\times \exp\left[-j2\pi((x_I - Mx_O)\tilde{x}_p + (y_I - My_O)\tilde{y}_p)\right]$$
$$\times d\tilde{x}_p d\tilde{y}_p U_O(x_O, y_O) dx_O dy_O.$$

That is, $U_I(x, y)$ and $U_O(x, y)$ have the following relationship:

$$U_I(x_i, y_i) = h(x_O, y_O) * \frac{1}{|M|}U_O(x_O, y_O), \quad (3)$$

where

$$h(x_O, y_O) = \iint_{-\infty}^{\infty} P(\lambda z_I \tilde{x}_p, \lambda z_I \tilde{y}_p)$$
$$\times \exp\left[j\frac{k}{2}\left(\frac{1}{z_O} + \frac{1}{z_I} - \frac{1}{F}\right)((\lambda z_I \tilde{x}_p)^2 + (\lambda z_I \tilde{y}_p)^2)\right] \quad (4)$$
$$\times \exp\left[-j2\pi(x_O \tilde{x}_p + y_O \tilde{y}_p)\right] d\tilde{x}_p d\tilde{y}_p$$
$$= \mathcal{F}\{P(\lambda z_I \tilde{x}_p, \lambda z_I \tilde{y}_p)W(z_i, \lambda z_I \tilde{x}_p, \lambda z_I \tilde{y}_p)\}$$

and

$$W(z_I, \lambda z_I \tilde{x}_p, \lambda z_I \tilde{y}_p)$$
$$= \exp\left[j\frac{k}{2}\left(\frac{1}{z_O} + \frac{1}{z_I} - \frac{1}{F}\right)((\lambda z_I \tilde{x}_p)^2 + (\lambda z_I \tilde{y}_p)^2)\right]. \quad (5)$$

Here, $h$ is the impulse response of the lens system, which is equal to the Fourier transform of the product of the pupil function $P$ and the aberration function $W$ for a given $z_O$.

Because the light emitted from the display is actually incoherent, we replace the complex fields $U_I(x, y)$ and $U_O(x, y)$ in Eq. (5) with intensities $I_I(x, y)$ and $I_O(x, y)$, respectively, and obtain

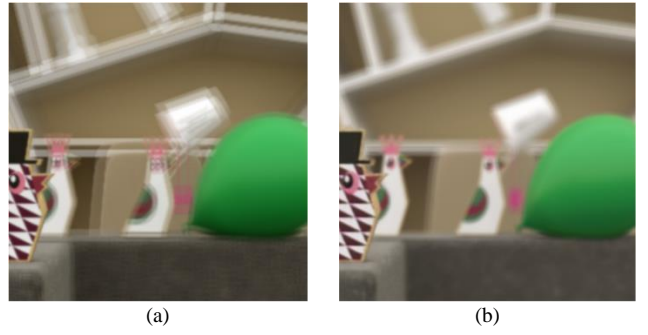

(a) (b)

Fig. 2. Example refocused images computed from the light fields generated by (a) TDM-LFD and (b) CTDM-LFD.

$$I_I(x_I, y_I) = |h(x_O, y_O)|^2 * I_O(\frac{x_O}{M}, \frac{y_O}{M}), \quad (6)$$

where constant factors are dropped for simplicity. Therefore, by assigning different values to $z_I$, we can compute the refocused images of $I_O$. We have thus derived the image formation formula for our optics system with one input image.

Next, we describe the formation of a refocused image from a light field consisting of a set of sub-aperture views. Without loss of generality, we consider a light field $L$ taken from a 2-D regular grid (2-D regular lattice) $G$ of viewpoints on the $(s, t)$-plane with sampling period $\Delta \mathbf{s} = (\Delta s, \Delta t)$. For each sub-aperture view $L_{\mathbf{s}}, \mathbf{s} \in G$, we represent the corresponding aperture by a shifted rectangular function

$$P_{\mathbf{s}}(\mathbf{x}) = \text{rect}(\frac{\mathbf{x} - \mathbf{s} \odot \Delta \mathbf{s}}{\text{aperture width}}), \quad (7)$$

where $\odot$ denotes element-wise multiplication. Let $h_{\mathbf{s}, z_I}$ denote the impulse response with pupil function $P_s$ and image distance $z_I$. Then the light field perceived at $z_I$, which is a refocused image on the retina, can be described by

$$R_{z_I} = \sum_{\mathbf{s} \in G}(|h_{\mathbf{s}, z_I}|^2 * L_{\mathbf{s}}). \quad (8)$$

The above derivation yields a simple way to build a light field display. However, the aperture allows only a small amount of light to go through because it is usually smaller than the light field angular sampling interval. Besides, aliasing occurs in the defocus regions due to low angular resolution, as shown in Fig. 2. These disadvantages of TDM-LFD motivated us to design coded time-division multiplexing for light field display.

IV. OPTIMIZING CTDM-LFD WITH DEEP LEARNING

Instead of using traditional gird-like apertures $P_{\mathbf{s}}$ with light field $L$ as the display panel input, we use specifically designed apertures $\hat{\mathbf{P}}$ with the light field encoded by a function $f$ as the display panel input, see the definition in Table II. We optimize $\hat{\mathbf{P}}$ and $f$ simultaneously with deep learning to improve the quality of the perceived light field. The CTDM-LFD pipeline is shown in Fig. 3 and the hyper-parameters of CTDM-LFD are shown in Table III. The configuration is shown in Fig. 1(d).

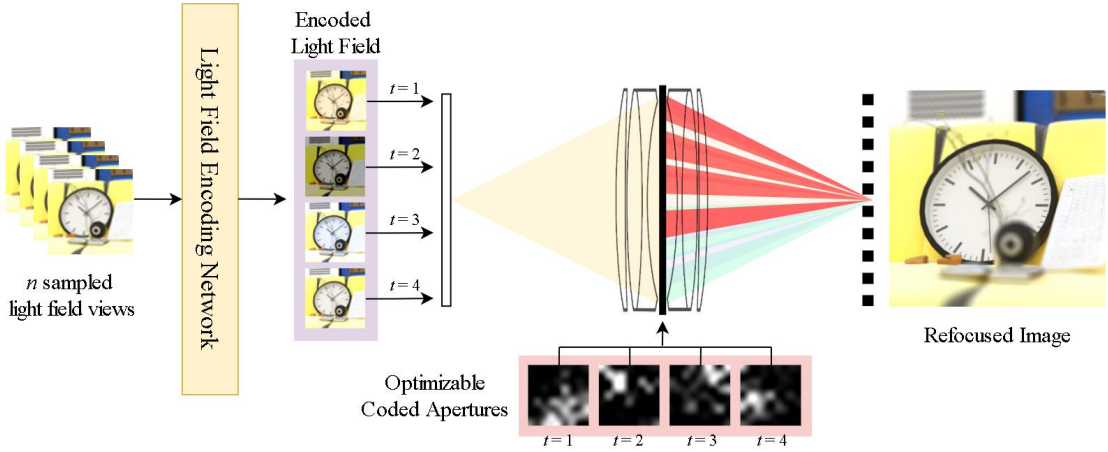

Fig. 3. Overview of the CTDM-LFD pipeline. Sampled light field views are first transformed by a light field encoding network into images that are more suitable for display. Then the encoded images pass through the optimized coded apertures and the lens system. Finally, the light field is projected to the human eye, where a refocused image is formed. The perception of the light field is effected through the persistence of vision. For the CTDM-LFD in this figure, $n = 4$ and $k = 4$.

In this section, we describe the details of the proposed CTDM-LFD and the modeling of its components by fully differentiable operations to facilitate the optimization of CTDM-LFD with deep learning.

### A. Light Field Encoding Network

We model $f$ with a convolutional neural network (CNN) and call it the light field encoding network.

Given $n$ sub-aperture views $V_1$, $V_2$, …, $V_n$ of a light field $L$, we concatenate them into a feature map $M_{in} \in \mathbb{R}^{H \times W \times 3n}$, where $H \times W$ is the spatial resolution of the light field $L$. $M_{in}$ is then input to our light field encoding network to generate $M_{out} \in \mathbb{R}^{H \times W \times 3k}$. We split $M_{out}$ by its last dimension into $k$ images $I_{c,1}$, $I_{c,2}$, …, $I_{c,k} \in \mathbb{R}^{H \times W \times 3}$. It should be noted that these images are not light field views, but rather encoded versions of them. We cannot supervise our model directly on the encoded views since we do not have their ground truth. As shown in Section IV, we optimize our model to fit ground truth focal stacks, and the model itself learns to generate optimal encoded light field views.

### B. Point Spread Function at Each Display Plane

In our CTDM-LFD, the point spread function of the display at a specific image distance $z_I$ is jointly determined by the coded aperture pattern and $z_I$. We denote this point spread function by

$$h_{\hat{P}_i, z_I} = \mathcal{F}\{\hat{P}_i(\lambda z_I \tilde{x}_p, \lambda z_I \tilde{y}_p) W(z_i, \lambda z_I \tilde{x}_p, \lambda z_I \tilde{y}_p)\}, \quad (9)$$

where $\hat{P}_i \in \mathbb{R}^{l \times l}$ is the $i^{th}$ coded aperture of the CTDM-LFD and $l$ denotes the *resolution* of the coded aperture. Note that each $\hat{P}_i$ is optimizable.

According to Eq. (6), as an encoded light field image $I_{c,i}$ passes through the coded aperture $\hat{P}_i$ and lens, the display generates an output $|h_{\hat{P}_i,d}|^2 * I_{c,i}$ at distance $d$ from the lens, where $*$ is the 2-D convolution operator.

### C. Merging Display Output with Refocused Images

For each light field $L$, the display panel sequentially displays the encoded light field images $I_{c,1}$, $I_{c,2}$, …, $I_{c,k}$. As the display panel switches to another image, the shutter also switches to the corresponding coded aperture in $\{\hat{P}_1, \hat{P}_2, …, \hat{P}_k\}$. Note that the set of apertures $\hat{P}$ is fixed for a CTDM-LFD. If the refresh rates of the display panel and shutter are high enough, the perception of $k$ consecutive images as a refocused image due to the persistence of vision can be described by

$$R_{z_I} = \sum_{t=1}^{k} |h_{\hat{P}_t, z_I}|^2 * I_{c,t} \in \mathbb{R}^{H \times W \times 3}. \quad (10)$$

### D. Obtaining the Focal Stack

Based on the above discussion, the perception of a light field $L$ at image distance $z_I$ is a refocused image $R_{z_I}$ on the retina. Since a viewer can focus at any image distance, the perceived light field is ultimately equivalent to a continuous focal stack. Our goal is to minimize the reconstruction error between this perceived focal stack and the ground truth focal stack.

However, rather than directly computing the continuous focal stack, we construct a discrete focal stack $F_c =$

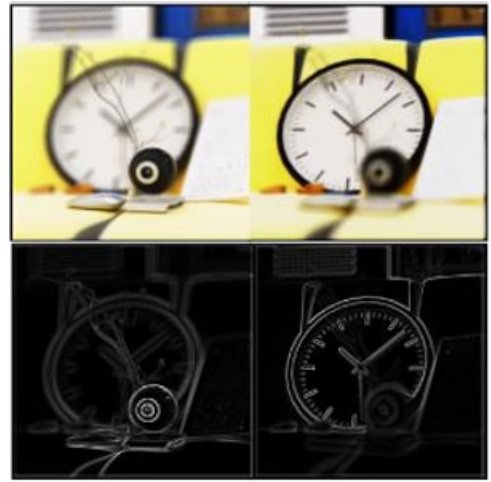

Fig. 4. Normalized focus measure $\bar{U}_j$ for two example refocused images. Notice that the regions with high focus measure shift from front to back, indicating the change of in-focus region.

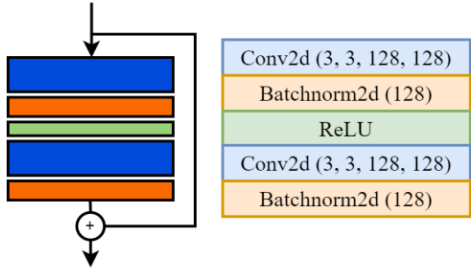

Fig. 5. A residual block in the light field encoding network. The input to the residual block is added to the final convolution block output.

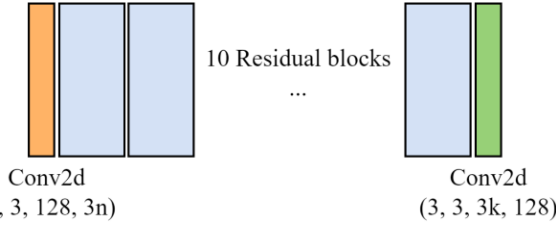

Fig. 6. The light field encoding network $f$ consists of an input 2-D convolutional layer, 10 residual blocks, and a final output 2-D convolutional layer. A 2-D convolution kernel is defined as (kernel_size, kernel_size, output channel, input channel).

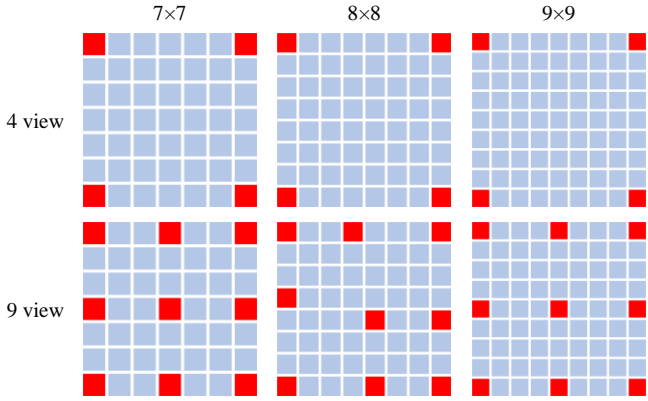

Fig. 7. Sub-sub-aperture views (red blocks) sampled from 7×7, 8×8, and 9×9 light fields. For the 8×8 case, we sample the light field such that the sampled views approximately form a 3×3 grid.

$\{R_{z_I,1}, R_{z_I,2}, ..., R_{z_I,m}\}$ that consists of $m$ images refocused at different depths to facilitate the optimization. Furthermore, the discrete focal stack computed from a full angular resolution light field using Eq. (8) is used as the ground truth $F_g$.

*E. Optimizing CTDM-LFD*

Merely optimizing the L1 distance

$$L = \| F_c - F_g \|_1 \quad (11)$$

between $F_g$ and $F_c$ means that the reconstruction loss has equal weight for all pixels. In our case, however, we want to emphasize the reconstruction of certain regions (e.g. in-focus or defocused regions) in a focal stack. Hence, we propose a *weighted reconstruction loss*:

$$L = \mathcal{W} \odot \| F_c - F_g \|_1, \quad (12)$$

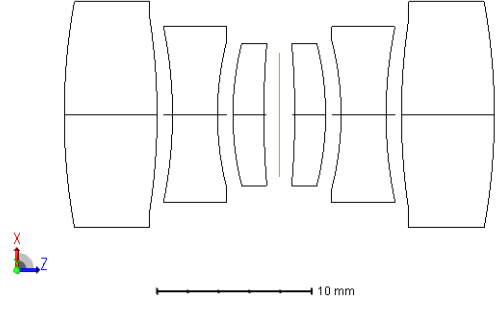

Fig. 8. Example optics design. The structure is symmetric, and the aperture is at the orange plane in the center.

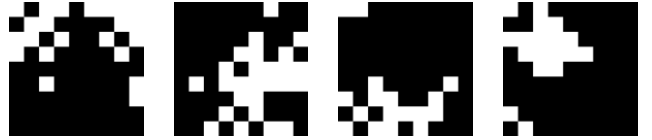

Fig. 9. The binary coded apertures used for Zemax optics simulation. The resolution of the coded apertures is 9×9 (that is, $l$=9).

TABLE IV
PARAMETERS OF THE FIRST THREE LENSES

| No | Type | R | T | Nd | Vd |
|---|---|---|---|---|---|
| 1 | SPH | 41.066 | 6.007 | 1.67790 | 50.72 |
| 2 | SPH | -41.115 | 1.000 | | |
| 3 | SPH | -27.799 | 2.912 | 1.61340 | 44.27 |
| 4 | SPH | 19.276 | 1.000 | | |
| 5 | SPH | 18.834 | 2.000 | 1.69350 | 50.81 |
| 6 | SPH | 54.311 | 1.000 | | |

where the $j^{th}$ weight map $\mathcal{W}_j \in \mathbb{R}^{H \times W}$, $1 \leq j \leq m$, in $\mathcal{W} \in \mathbb{R}^{m \times H \times W}$ is the weight map that specifies the reconstruction weight for each pixel in the $j^{th}$ refocused image. To obtain $\mathcal{W}_j$, we first compute the *focus measure* $U_j \in \mathbb{R}^{H \times W}$ from the $j^{th}$ refocused image $R_{z_I,j}$ by

$$U_j = \| \nabla_x R_{z_I,j} \|_1 + \| \nabla_y R_{z_I,j} \|_1, \quad (13)$$

which measures the level of focus of each pixel in $R_{z_I,j}$, as shown in Fig. 4.

Then, we compute $\mathcal{W}_j$ by taking the exponential of the normalized and scaled $U_j$,

$$\mathcal{W}_j = e^{\beta \cdot \bar{U}_j}, \quad (14)$$

where

$$\bar{U}_j = \frac{U_j - \min U_j}{\max U_j - \min U_j}. \quad (15)$$

A large value of $\beta$ indicates that the model emphasizes the reconstruction of in-focus pixels. For example, $\mathcal{W}_j$ is simply a matrix of ones when $\beta = 0$, meaning that every pixel has the same reconstruction weight.

TABLE V
PERFORMANCE OF CTDM-LFD ON DIFFERENT DATASETS

| Method | PSNR / SSIM / LPIPS | | |
| --- | --- | --- | --- |
| | INRIA-syn | INRIA-real | SIGGRAPH16 |
| Baseline ($n = 4$) | 28.56/0.8485/0.2626 | 26.39/0.8932/0.1434 | 32.92/0.9285/0.1199 |
| CTDM-LFD ($n = 4$, $k = 4$) | **32.29/0.9374/0.1284** | **30.21/0.9602/0.0736** | **37.57/0.9705/0.03119** |
| Baseline ($n = 9$) | 34.73/0.9688/0.1385 | 36.32/0.9872/0.0363 | 40.61/**0.9858**/0.0397 |
| CTDM-LFD ($n = 9$, $k = 8$) | **36.41/0.9858/0.0632** | **39.97/0.9891/0.0186** | **41.21**/0.9739/**0.0107** |

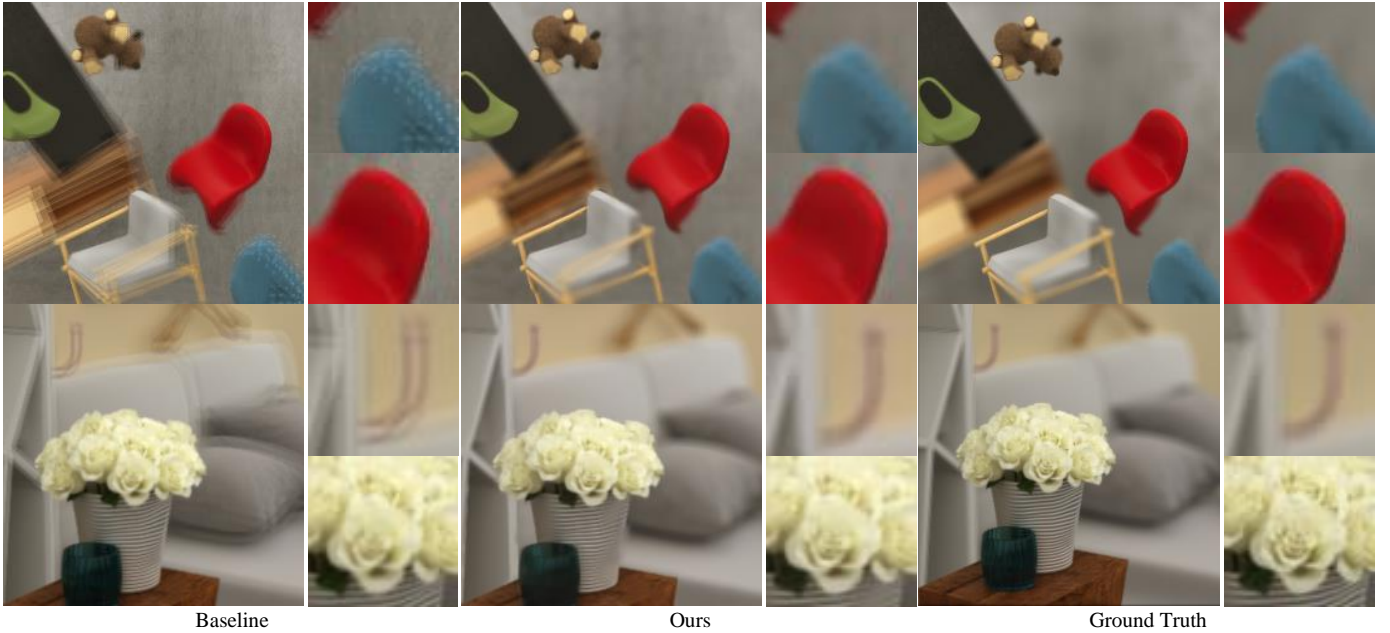

Baseline            Ours            Ground Truth

Fig. 10. Visual quality comparison (n = 4) on the INRIA Synthetic Dataset. The figure shows refocused images computed from the light fields generated by CTDM-LFD and TDM-LFD, and the ground truth image. The refocused image computed from the light field generated by TDM-LFD has severe aliasing in defocus regions due to undersampling. As shown in the zoomed-in image patches, our method preserves the sharpness of in-focus regions and has natural blur in defocus regions.

## V. EXPERIMENTAL SETUP AND RESULTS

In this section, we perform qualitative and quantitative comparisons between our CTDM-LFD and the baseline TDM-LFD on two light field datasets. Furthermore, we verify our technique using the optical design software Zemax by implementing an example design of our proposed CTDM-LFD. Finally, ablation studies are conducted on various configurations of the display optimization pipeline.

### A. Datasets

We verify the effectiveness of our method on three publicly available light field datasets: the INRIA synthetic light field dataset (denoted by *INRIA-syn*) [53], the INRIA Lytro dataset [54] (denoted by *INRIA-real*), and the dataset used in the paper [55] (denoted by *SIGGRAPH16*).

*INRIA-syn* dataset consists of 39 light field scenes with 9×9 angular resolution; *INRIA-real* dataset consists of 63 light field scenes with 7 × 7 angular resolution; *SIGGRAPH16* consists of 130 light field scenes with 8 × 8 angular resolution. We split *INRIA-syn, INRIA-real,* and *SIGGRAPH16* datasets into training/testing data and obtain 31/8, 51/12, and 100/30 light fields, respectively.

### B. Evaluation Metrics

We use three image quality assessment metrics, peak signal-to-noise ratio (PSNR), structural similarity index measure (SSIM), and learned perceptual image patch similarity (LPIPS) to evaluate the quality of the light fields displayed using our pipeline.

### C. Training Details

The light field encoding network $f$ consists of ten stacked residual blocks, as detailed in Fig. 5 and Fig. 6. We clamp the network output to the range [0, 1] with the *sigmoid* function. For the results shown in Sections V.D and V.E, we choose $\beta = 2$ for the weighted reconstruction loss.

All networks are trained end-to-end from scratch for 10000 epochs. We use the Adam optimization algorithm with default parameters $\beta_1 = 0.9$, $\beta_2 = 0.999$, $\epsilon = 1e-8$, and learning rate 0.001.

### D. Quantitative Results

To evaluate the performance of our light field display, we compute a focal stack from the generated light field. The quality of the focal stack of CTDM-LFD is compared to that of the baseline TDM-LFD. Four and nine sub-aperture views are used in this experiment for light field. Note again that the quality of a light field is evaluated on the focal stack computed from the light field, not on a single refocused image.

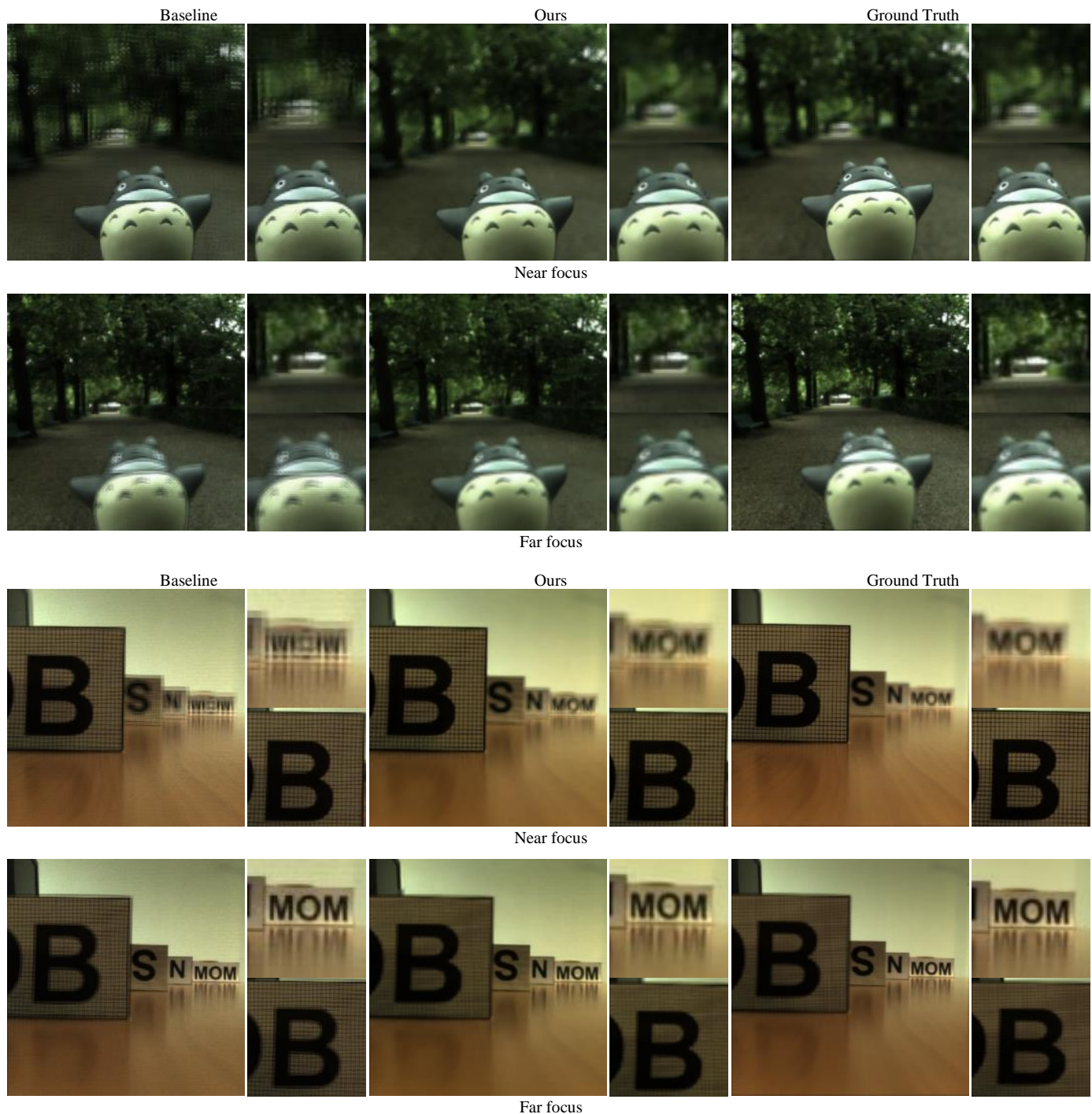

Fig. 11. Visual quality comparison (n = 4, k=4) on the INRIA Lytro Dataset. The figure shows the refocused images computed from the light fields generated by CTDM-LFD and TDM-LFD. The focal stack computed from the light field generated by the TDM-LFD has severe aliasing in defocus regions due to undersampling. Our method preserves the sharpness of in-focus regions and has natural blur in defocus regions.

For the four-view setting, the sampled light field views are the four corner views. For the nine-view setting, the sampled light field views form a 3×3 grid in the original light field, as shown in Fig. 7.

For the four-view and nine-view settings in our method, we consider CTDM-LFD with $k = 4$ and $k = 8$, respectively.

PSNR, SSIM, and LPIPS scores under these two settings are summarized in Table V. We can see that focal stacks computed from the light field generated by the CTDM-LFD have better PSNR, SSIM, and LPIPS scores than the baseline TDM-LFD for $n = 4$ and $n = 9$ on all datasets in all cases except for n = 9 on the *SIGGRAPH16* dataset in terms of the SSIM metric. Note that we use 4/9 sub-aperture views only to ensure fair comparison between our CTDM and baseline SDM designs, since SDM designs divide the display panel into $n^2$ patches ($n = 2,3$ in this case). If we use all light field sub-aperture views to generate the encoded light field images, the performance of CTDM-LFD could be better.

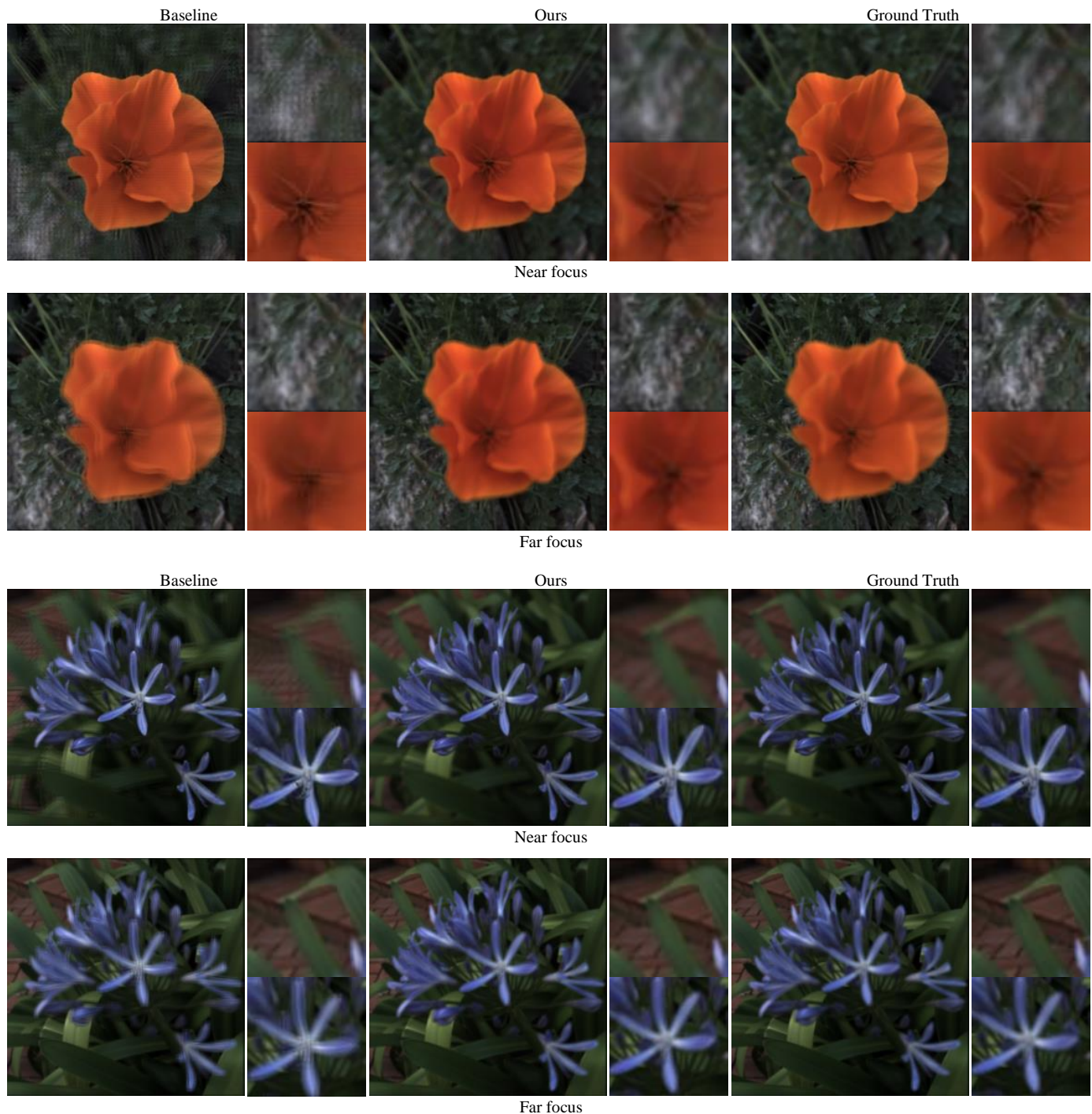

Fig. 12. Visual quality comparison (n = 4, k=4) on the SIGGRAPH16 Dataset. The figure shows the refocused images computed from the light fields generated by CTDM-LFD and TDM-LFD. The focal stack computed from the light field generated by the TDM-LFD has severe aliasing in defocus regions due to undersampling. Our method preserves the sharpness of in-focus regions and has natural blur in defocus regions.

### E. Qualitative Results

We also perform visual quality comparisons of refocused images sampled from the focal stacks computed from the light fields generated by our CTDM-LFD and the baseline TDM-LFD. From Fig. 10, Fig. 11, and Fig. 12, severe aliasing in the defocused regions of the refocused images of TDM-LFD can be observed. On the contrary, our CTDM-LFD method preserves the sharpness of in-focus regions and has natural blur in defocused regions.

Fig. 17 shows the full focal stacks (nine focal planes sampled for each focal stack) computed from the light fields generated by our CTDM-LFD. We can see that the in-focus regions gradually shift from the front to the back of the light field scene. The infocus regions remain sharp while defocus regions have a natural defocus blur.

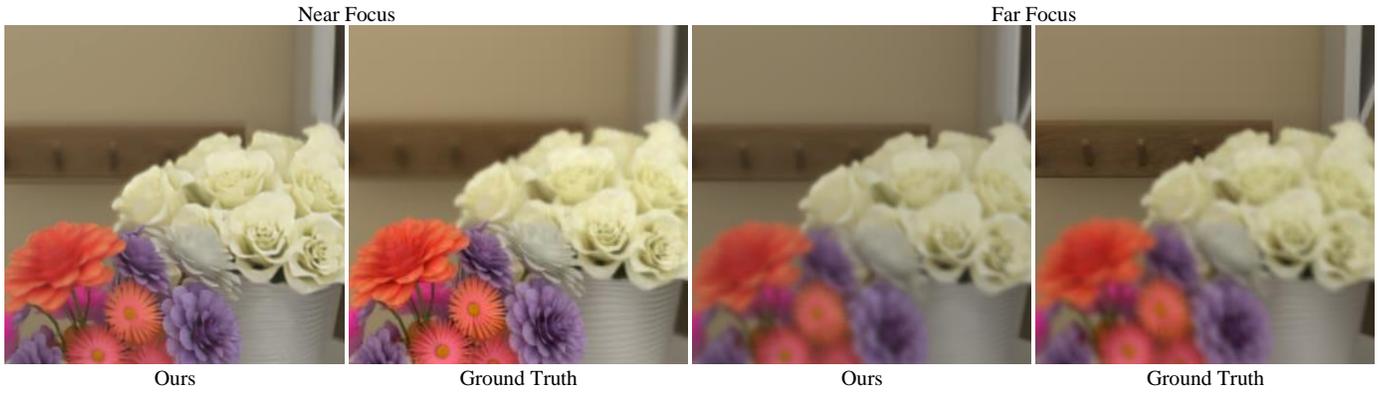

Fig. 13. Refocused images computed from the light field generated by CTDM-LFD using optics design software.

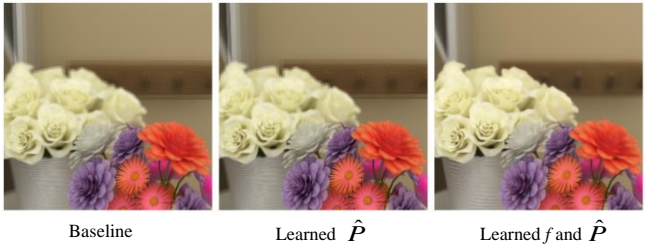

Fig. 14. Visual quality comparisons of refocused images computed from the light field generated by the baseline TDM-LFD (left), the CTDM-LFD with learned coded apertures (middle), and the CTDM-LFD with learned coded apertures and encoded light field (right). Notice that the severe aliasing found in the left and middle images is mitigated in the right image.

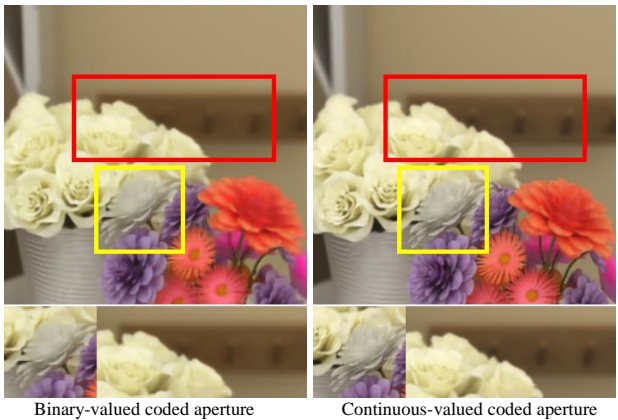

Fig. 15. Visual quality comparison of refocused images computed from the light field generated by the CTDM-LFD using binary- and continuous-valued apertures.

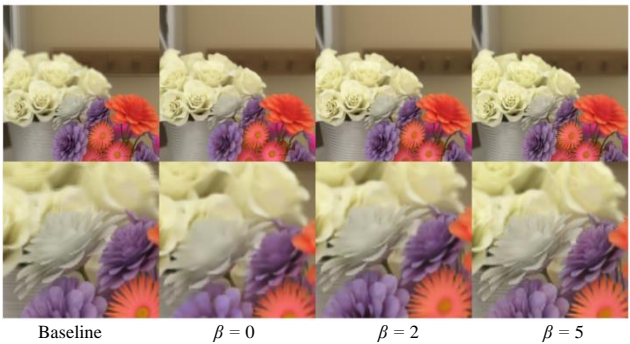

Fig. 16. Visual comparisons of refocused images computed from the light field generated by the CTDM-LFD with different $\beta$ values.

### F. Optics Simulation

To verify the practicability of our CTDM-LFD technique, we use the optics design software Zemax to design an example of CTDM-LFD and perform optical simulation. As shown in Fig. 8, we use a unit magnification relay lens in the experiment. The whole system is easy to manufacture because it consists of only spherical surfaces and common glass materials (Ohara S-NBM51, S-LAL56, and LAL58), and the detailed parameters are shown in Table IV. The lens system is symmetric and a binary-valued aperture is placed at the center. (Because it is difficult to simulate continuous-valued apertures in Zemax, we only test binary-valued apertures in the experiment. The performance of a CTDM-LFD with binary-valued apertures is slightly lower than a CTDM-LFD with continuous-valued ones, which will be discussed in Section V.G.) The optimized 9×9 binary aperture pattern is shown in Fig. 9.

Note that the proposed CTDM-LFD is not limited to the relay lens. For most imaging or projector systems, we can similarly put the aperture at their Fourier plane.

The optics simulation result is shown in Fig. 13. We can see that our CTDM-LFD displays a light field that can be focused at different depths. Furthermore, the bokeh in the refocused images are much more natural compared to that of TDM-LFD.

### G. Ablation Studies

We perform ablation studies on various design variables (hyper-parameters) in our CTDM-LFD and show performance comparisons between different design choices. All experiments are conducted on the INRIA Synthetic Light Field Dataset.

1) *Learned f and $\hat{P}$*

In the baseline TDM-LFD shown in Fig. 1(c), $f$ is simply the identity function. We compare the performance of three cases: baseline, only $\hat{P}$ learned, and both $f$ and $\hat{P}$ learned. We choose $n = 4$, $k = 4$, $l = 9$ in this experiment. From Fig. 14 and Table VI, we can see that merely optimizing $\hat{P}$ leads to performance improvement by over 1 dB. Also optimizing the light field encoding mapping $f$ (which is our final CTDM-LFD setting) further improves the performance by nearly 3 dB.

2) *Value of coded apertures*

The quantization of the values of coded apertures can be specified by the user depending on the application. For example, if the coded aperture is implemented by a gray-scale liquid

TABLE VI
PERFORMANCE OF CTDM-LFD WITH DIFFERENT LEARNED COMPONENTS

| Method | PSNR | SSIM | LPIPS |
|---|---|---|---|
| Baseline | 28.56 | 0.8485 | 0.2626 |
| CTDM-LFD with *Learned* $\hat{P}$ | 29.85 | 0.8762 | 0.2277 |
| CTDM-LFD with *Learned f and* $\hat{P}$ | **32.29** | **0.9374** | **0.1284** |

* Setting: $n = 4$, $k = 4$, $l = 9$, continuous coded aperture.

TABLE VII
PERFORMANCE OF CTDM-LFD WITH BINARY- AND CONTINUOUS-VALUED CODED APERTURES

| Method | PSNR | SSIM | LPIPS |
|---|---|---|---|
| *Binary-valued* | 31.24 | 0.9249 | 0.1415 |
| *Continuous-valued* | **32.29** | **0.9374** | **0.1284** |

* Setting: $n = 4$, $k = 4$, $l = 9$

TABLE VIII
PERFORMANCE OF CTDM-LFD WITH SYMMETRIC AND FREE-VALUED CODED APERTURES

| Method | PSNR | SSIM | LPIPS |
|---|---|---|---|
| *Symmetric* | 31.06 | 0.9283 | 0.1413 |
| *Free-valued* | **32.29** | **0.9374** | **0.1284** |

* Setting: $n = 4$, $k = 4$, $l = 9$

TABLE IX
PERFORMANCE OF CTDM-LFD WITH DIFFERENT $n$ AND $k$ VALUES

| $n$ | PSNR/SSIM/LPIPS | | |
|---|---|---|---|
| | $k = 4$ | $k = 8$ | $k = 16$ |
| 4 | 32.29/**0.9374/0.1284** | **33.12**/0.9304/0.1428 | 32.43/0.9341/0.1284 |
| 9 | 33.27/0.9723/0.0759 | **36.41/0.9858/0.0632** | 34.15/0.9795/0.0693 |

* Setting: $l = 9$

TABLE X
PERFORMANCE OF CTDM-LFD WITH DIFFERENT $\beta$ VALUES

| $\beta$ | PSNR | SSIM | LPIPS |
|---|---|---|---|
| 0 (equal weighting) | 31.25 | 0.9245 | 0.1561 |
| 2 | **32.29** | 0.9374 | 0.1284 |
| 5 | 32.07 | **0.9376** | **0.1248** |

* Setting: $n = 4$, $k = 4$, $l = 9$

crystal on silicon (LCoS), we can optimize the model with continuous-valued apertures. On the other hand, if the coded aperture is implemented by a binary liquid crystal array, the model can be optimized with binary-valued apertures.

We test two quantization configurations of our coded apertures: *continuous-valued* and *binary-valued.* For the binary-valued apertures, we use the following modified training and inference procedure. During training, $\hat{P}$ is passed through the following sigmoid function with large temperature value $t$ before computing the point spread functions:

$$S(x) = \frac{1}{1+e^{-tx}} \quad (15)$$

During inference, $C$ is binarized to $\{0, 1\}$ where values larger than 0.5 is replaced with 1 and values less than 0.5 are replaced with 0.

From Table VII and Fig. 15, we can see that CTDM-LFD with continuous apertures has better performance. This is expected since the possible patterns that continuous coded apertures can have are almost infinite, which are more flexible for optimization.

3) *Symmetry of coded apertures*

Symmetry is sometimes a desirable property in traditional optics design. In our pipeline, we experiment on this concept in the $k = 4$ setting by restricting the three other coded apertures to be mirrored versions of the first one.

Table VIII shows that symmetry of coded apertures would degrade the quality of refocused images. One possible reason is that the capacity of the model is reduced due to the reduced number of optimizable parameters.

4) *Different values of n and k*

We test different values of $n$ (the number of sampled light field sub-aperture views) and $k$ (the number of encoded light field images) and evaluate the quality of light fields displayed by CTDM-LFD. We choose $l = 9$ in this experiment.

From Table IX, we can see that larger $n$ has better performance due to the increase of light field information. However, larger $k$ does not necessarily lead to better performance. In fact, $k$ should not be too large in practice, since the required frame rate of CTDM-LFD increases linearly with $k$. Generally speaking, large $k$ results in a lower refresh rate for the display. Hence, if image quality is preferred over refresh rate, a large $k$ is better. Small $k$, on the other hand, results in a high refresh rate but low image quality

5) *Weighted reconstruction loss parameter β:*

The $\beta$ parameter in the weighted reconstruction loss controls the relative weighting of reconstruction loss between in-focus and defocus regions in the focal stack. We compare the performance of the CTDM-LFD with different $\beta$ values.

We find that global image quality metrics cannot fully reflect the quality of the focal stack. From Table X and Fig. 16, the PSNR is highest for $\beta = 2$, yet details in in-focus regions are more preserved for $\beta = 5$. This means that $\beta$ is a hyper-parameter that users can tune to fit a particular application, depending on how much the in-focus regions should be emphasized.

## VI. CONCLUSION

In this paper, we propose a novel display technique called coded time-division light field display (CTDM-LFD), which projects encoded light field sub-aperture views to a viewer's eyes and offers correct cues for vergence-accommodation reflex. By jointly optimizing the light field encoding network and coded aperture pattern, our pipeline can display high-quality light fields from a sparse light field with minimal aliasing. By simulating light transport and image formation with Fourier optics, we can learn the encoding network and coded aperture pattern via deep learning in an end-to-end fashion. Theoretical and optical simulation results show that our method can display light fields with higher quantitative and qualitative quality than baseline display designs.

ACKNOWLEDGEMENT

The authors thank Dr. C.K. Lee of NVIDIA AI Technology Center (NVAITC) for valuable discussions and NVAITC for providing GPU computing resources.

REFERENCES

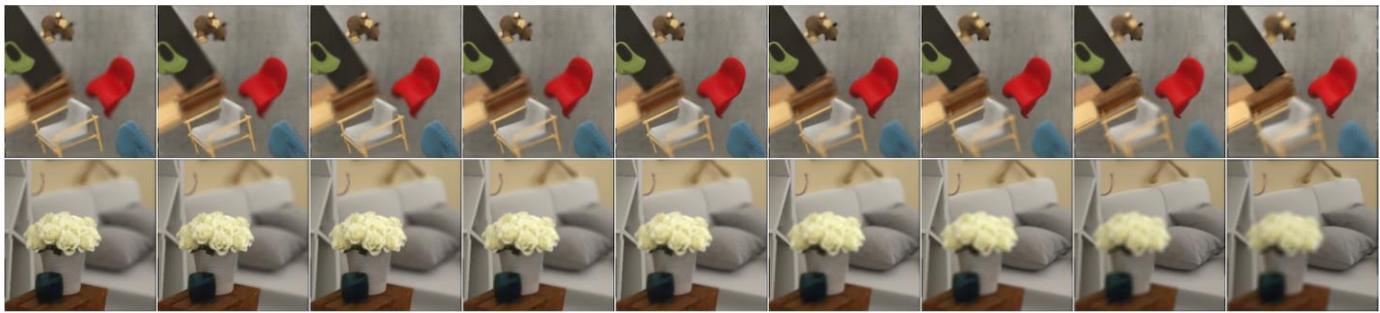

(a) INRIA Synthetic Dataset Continuous Focal Stack Result

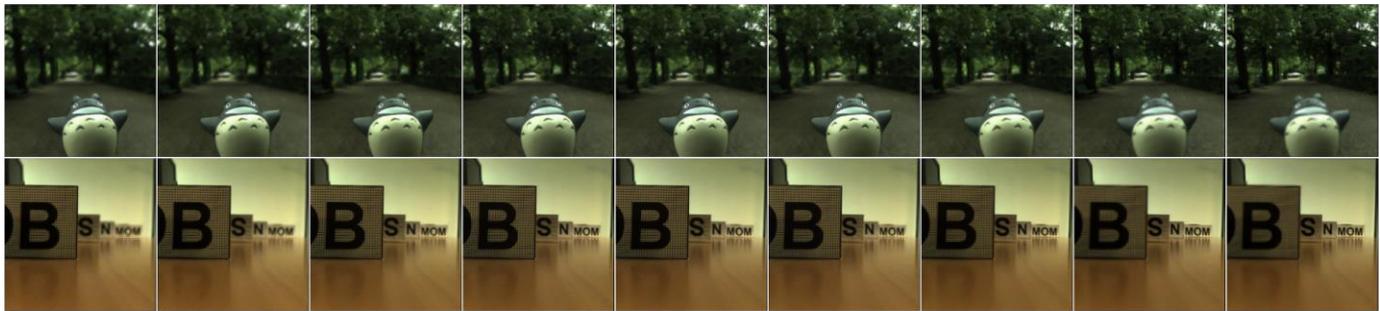

(b) INRIA Lytro Continuous Focal Stack Result

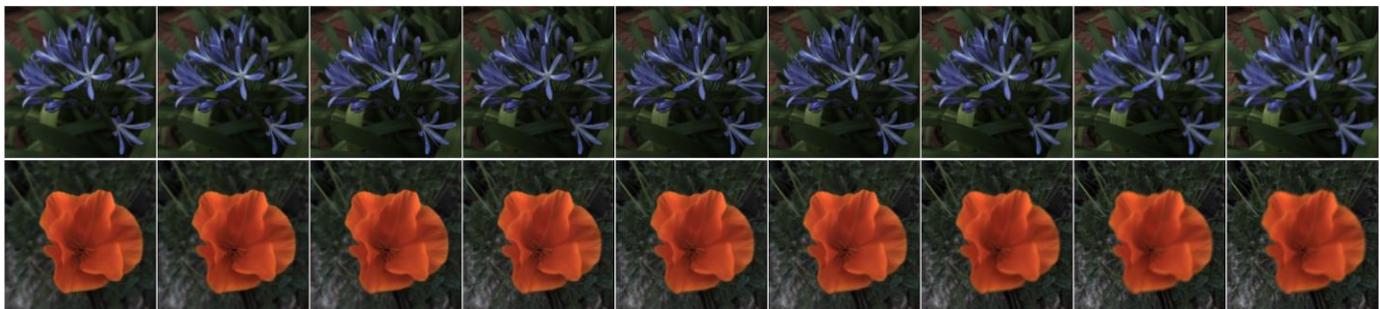

(c) Stanford Dataset Continuous Focal Stack Result

Fig. 17. Continuous focal stacks computed from the light fields generated by CTDM-LFD (n=4, k=4). Viewing from left to right, we can see that the in-focus regions gradually and smoothly moves from the front to the back of the scene.

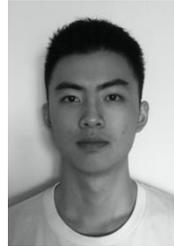
**Chung-Hao Chao** is currently working toward a B.S degree in electrical engineering at National Taiwan University, from 2017. He was an undergraduate research assistant at Visual and Learning Lab (VLL), National Taiwan University. He is currently a research assistant at Multimedia Processing and Communications Lab (MPAC) at National Taiwan University and intern at NVIDIA AI Technology Center (NVAITC). His research interests include image processing, light field processing, and computational photography and displays.

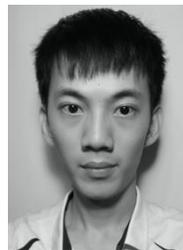
**Chang-Le Liu** is currently a working toward a M.S degree in communication engineering at National Taiwan University, from 2020. He was a summer intern as a research assistant with the Research Center for Information Technology Innovation, Academia Sinica, Taipei, Taiwan, and was involved in research in speech enhancement. His current research topic includes image and audio signal processing and light field photography and rendering.

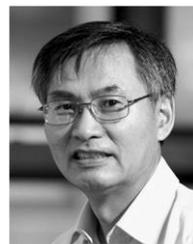
**Homer H. Chen** (S'83–M'86–SM'01–F'03) received the Ph.D. degree in electrical and computer engineering from the University of Illinois at Urbana–Champaign. His professional career has spanned industry and academia. Since August 2003, he has been with the College of Electrical Engineering and Computer Science, National Taiwan University, where he is a Distinguished Professor. Prior to that, he held various research and development management and engineering positions at U.S. companies over a period of 17 years, including AT&T Bell Labs, Rockwell Science Center, iVast, and Digital Island. He was a U.S. delegate for ISO and ITU standards committees and contributed to the development of many new interactive multimedia technologies that are now a part of the MPEG-4 and JPEG-2000 standards. His professional interests lie in the broad area of multimedia signal processing and communications.

Dr. Chen serves on the IEEE Signal Processing Society Awards Board and the Senior Editorial Board of the IEEE Journal on Selected Topics in Signal Processing. He was a Distinguished Lecturer of the IEEE Circuits and Systems Society from 2012 to 2013. He served on the IEEE Signal Processing Society Fourier Award Committee and the Fellow Reference Committee from 2015 to 2017. He was a General Chair of the 2019 IEEE International Conference on Image Processing. He was an Associate Editor of the IEEE Transactions on Circuits and Systems for Video Technology from 2004 to 2010, the IEEE Transactions on Image Processing from 1992 to 1994, and Pattern Recognition from 1989 to 1999. He served as a Guest Editor for the IEEE Transactions on Circuits and Systems for Video Technology in 1999, the IEEE Transactions on Multimedia in 2011, the IEEE Journal of Selected Topics in Signal Processing in 2014, and Multimedia Tools and Applications (Springer) in 2015.